\documentclass[twocolumn,preprintnumbers, amssymb,amsmath,aps,prd,nofootinbib,superscriptaddress,showpacs]{revtex4-1}

\usepackage{amsmath,amsfonts}
\usepackage{algorithmic}
\usepackage{algorithm}
\usepackage{array}
\usepackage{textcomp}
\usepackage{stfloats}
\usepackage{url}
\usepackage{verbatim}
\usepackage{graphicx}
\usepackage{dcolumn}
\usepackage{amsmath}
\usepackage{amsfonts}
\usepackage{mathrsfs}
\usepackage{physics}
\usepackage{bm}
\usepackage{dsfont}
\usepackage{enumerate}
\usepackage{slashed}
\usepackage[dvipsnames]{xcolor}
\usepackage{tikz, tikz-feynman}
\usepackage{pgfplots}
\pgfplotsset{compat=1.16}

\usepackage{epsfig}
\usepackage{bm}
\usepackage{amssymb}
\usepackage{amsmath}
\usepackage{color}
\usepackage{subfigure}
\usepackage[colorlinks,
            linkcolor=blue,
            anchorcolor=black,
            citecolor=blue
            ]{hyperref}
\usepackage{soul}

\begin{document}

\title{Studying Maximal Entanglement and Bell Nonlocality at an Electron-Ion Collider}

\author{Wei Qi}\email{qi.673@osu.edu}
\affiliation{School of Science and Engineering, The Chinese University of Hong Kong (Shenzhen), Longgang, Shenzhen, Guangdong, 518172, P.R. China}
\affiliation{Department of Physics, The Ohio State University, Columbus, OH 43210, USA}

\author{Zijing Guo}\email{zijingguo@link.cuhk.edu.cn}
\affiliation{School of Science and Engineering, The Chinese University of Hong Kong (Shenzhen), Longgang, Shenzhen, Guangdong, 518172, P.R. China}

\author{Bo-Wen Xiao}\email{xiaobowen@cuhk.edu.cn} 
\affiliation{School of Science and Engineering, The Chinese University of Hong Kong (Shenzhen), Longgang, Shenzhen, Guangdong, 518172, P.R. China}
\affiliation{Southern Center for Nuclear-Science Theory (SCNT), Institute of Modern Physics, Chinese Academy of Sciences, Huizhou 516000, Guangdong Province, China}

\begin{abstract}
In this paper, we propose to test quantum entanglement and Bell nonlocality at an Electron-Ion Collider (EIC). By computing the spin correlations in quark-antiquark pairs produced via photon-gluon fusion, we find that longitudinally polarized photons produce maximal entanglement at leading order, while transversely polarized photons generate significant entanglement near the threshold and in the ultra-relativistic regime. Compared to hadron colliders, the EIC provides a cleaner experimental environment for measuring entanglement through the $\gamma^\ast g \to q\bar{q}$ channel, offering a strong signal and a promising avenue to verify Bell nonlocality. This study extends entanglement measurements to the EIC, presenting new opportunities to explore the interplay of quantum information phenomena and hadronic physics in the EIC era.
\end{abstract}
\maketitle

\noindent \textit{Introduction.---}
As the quintessential phenomenon of quantum mechanics, quantum entanglement has historically been pivotal in establishing the probabilistic interpretation of quantum theory and revealing the genuinely non-local nature of quantum mechanics.

In 1935, the famous Einstein-Podolsky-Rosen (EPR) paradox \cite{Einstein:1935rr} was proposed as a challenge to the concept of entanglement and, more broadly, the orthodox Copenhagen interpretation of quantum mechanics. The EPR paradox suggested that quantum mechanics might be incomplete and raised the possibility of alternative interpretations based on hidden variables. Hidden variable theories aim to provide additional information, offering a complete, deterministic description of physical systems without relying on probabilistic concepts. For many years, it was believed that hidden variable theories and quantum mechanics would yield identical physical predictions, making it impossible to experimentally distinguish between them. In 1964, Bell ingeniously demonstrated that any local hidden variable theory is fundamentally incompatible with quantum mechanics through what is now known as the Bell inequality \cite{Bell:1964kc}. A few years later, Clauser, Horne, Shimony, and Holt \cite{Clauser:1969ny, Clauser:1974tg} proposed the Clauser-Horne-Shimony-Holt (CHSH) inequality, which generalized the Bell inequality and laid the foundation for a practical framework for experimental tests at the atomic level. Their work not only provided decisive evidence supporting quantum mechanics and rejecting local hidden variable theories but also paved the way for the development of the field of quantum information theory (see \cite{Plenio:2007zz, Wootters:1997id, Cerf:1996nb, Acin:2002zz, Cirelson:1980ry, Horodecki:1995nsk}, among many other references).

Recently, the study of quantum entanglement in high-energy collisions 
\cite{
Tu:2019ouv, Datta:2024hpn, Hentschinski:2023izh, Florio:2023dke,
Uwer:2004vp, Afik:2022dgh, Afik:2022kwm, Afik:2020onf, CMS:2023gjz, Cheng:2024btk, Aguilar-Saavedra:2024whi,
Chen:2013epa, Wu:2024mtj, Ashby-Pickering:2022umy,
Barr:2024djo, Ehataht:2023zzt, Guo:2024jch, Morales:2023gow,
Parke:1996pr, Ruzi:2024cbt, Aguilar-Saavedra:2022mpg, Altakach:2022ywa, Aoude:2024xpx,
Bernal:2023jba, Brandenburg:2002xr, Heinrich:2022idm,
Hentschinski:2024gaa, James:2001klt, Kowalska:2024kbs, Li:2023gkh,
Ma:2023yvd, Maltoni:2024csn, Sakurai:2023nsc, Severi:2022qjy, 
Subba:2024mnl, Severi:2021cnj,Wu:2024asu,Kharzeev:2017qzs,Aguilar-Saavedra:2024fig, Grabarczyk:2024wnk, Fabbrichesi:2023idl, Fabbrichesi:2024rec,
Li:2024luk, Shi:2004yt, Fabbri:2023ncz, Bramon:2001tb, Wu:2024ovc, Mahlon:1995zn, Dong:2023xiw, Aguilar-Saavedra:2022wam,
Aguilar-Saavedra:2022uye, Bernal:2023ruk, Bernal:2024xhm, Bi:2023uop, Fabbrichesi:2024wcd,
Fabbrichesi:2023cev, Ghosh:2023rpj, Grabarczyk:2024wnk,White:2024nuc,Han:2023fci, Du:2024sly,
Cheng:2025cuv, Morales:2024jhj,Yang:2024kjn}
 has attracted significant attention, and new opportunities to explore this fascinating phenomenon beyond the atomic scale have emerged.  
Both experimentalists and theorists have been exploring the production of top-antitop quark pair at the Large Hadron Collider (LHC), carrying out a meticulous experimental measurement \cite{CMS:2019nrx, ATLAS:2019zrq,ATLAS:2023fsd, CMS:2024pts} of entanglement by studying the correlation between their weak decay products\cite{Aguilar-Saavedra:2023hss, Aguilar-Saavedra:2024hwd}. In particular, the ATLAS and CMS Collaborations~\cite{ATLAS:2023fsd, CMS:2024pts} at the LHC have observed quantum entanglement between top quark pairs in proton-proton collisions at $\sqrt{s} = 13$~TeV. Typically, top quark pairs are produced via quark annihilation or gluon-gluon fusion channels. By measuring the angular correlations of the lepton pairs in top's weak decay products, one can access the transferred spin information. However, due to the contribution from the quark annihilation channel\cite{Severi:2021cnj,Fabbrichesi:2025psr}, it remains challenging to observe violations of Bell inequality or Bell nonlocality at the LHC.

As summarized in a recent article~\cite{Afik:2025ejh}, there have been many studies positioned at the interplay between quantum information theory and high-energy physics, examining various channels in electron-positron annihilation processes at dedicated colliders and in proton-proton collisions at the LHC. However, theoretical studies of electron-proton collisions are still lacking, and complementary measurements at the upcoming Electron-Ion Collider (EIC)~\cite{Accardi:2012qut,AbdulKhalek:2021gbh,Anderle:2021wcy} could potentially become significant. We find that the quark pair produced by a longitudinal virtual photon is always maximally entangled at leading order (LO), and the pair produced by a transverse photon is in a maximally entangled spin-singlet state near the threshold. This makes EICs ideal machines for measuring entanglement and Bell nonlocality.

This paper is structured as follows: we first introduce the general formalism for the density matrix of a two-qubit system and the criteria for quantum entanglement, then compute the spin density matrix for the quark-antiquark pair in the photon-gluon fusion process. Additionally, we present results for both longitudinal and transverse incident photons separately and demonstrate that entanglement and Bell nonlocality can be measured at the EIC. Finally, we comment on the experimental implications and possible future extensions of this work.

\noindent \textit{General formalism.---} Without loss of generality, any 2-qubit system (such as a quark-antiquark pair) can be represented by a $4 \times 4$ spin density matrix as follows
\begin{equation}        
       \rho = \frac{1}{4}( \mathds 1_4 + 
       B^+_i \sigma^i \otimes \mathds 1_2 +
       B^-_j \mathds 1_2 \otimes \sigma^j +
       C_{ij} \sigma^i \otimes \sigma^j ),  \label{rho}
\end{equation}
where $B^+_i=\tr[\rho (\sigma^i \otimes \mathds 1_2)]$ and $ B^-_j=\tr[\rho (\mathds 1_2 \otimes \sigma^j)]$ are the spin polarizations of each qubit, and $C_{ij} = \tr[\rho (\sigma^i \otimes \sigma^j )]$ is a measure of the spin correlation between qubits. For example, $C_{ij} = -\delta_{ij}$ indicates that the two spins are always anti-parallel and therefore form a spin singlet configuration $\ket{\Psi^-} = \frac{1}{\sqrt{2}}(\ket{\uparrow \downarrow} - \ket{ \downarrow \uparrow })$. Similarly, correlation matrices with $C= \text{diag}(1,1,-1), \text{diag}(1,-1,1), \text{diag}(-1,1,1)$ correspond to the spin triplet states and the other three Bell states, i.e., $\ket{\Psi^+} = \frac{1}{\sqrt{2}}(\ket{\uparrow \downarrow} + \ket{\downarrow \uparrow})$ and $\ket{\Phi^\pm} = \frac{1}{\sqrt{2}}(\ket{\uparrow \uparrow} \pm \ket{\downarrow \downarrow})$.

\begin{figure}[H]
\includegraphics[width=0.85\linewidth]{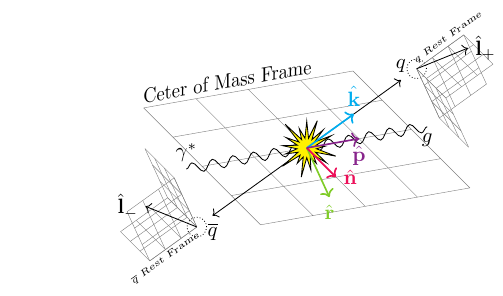}
\caption{\label{fig:frames} Illustration of the helicity basis vectors ${\hat{\vb n}, \hat{\vb r}, \hat{\vb k}}$ and the unit vector $\hat{\vb p}$ of the photon momentum, defined in the center-of-mass frame, alongside the unit vectors $\hat{\vb l}_+$, $\hat{\vb l}_-$ of the fermionic decay products (indicating momentum directions), shown in their respective parent particle rest frames, where the spin of each particle and the decay spin density matrices are also conventionally defined. Note that $\hat{\vb l}_+$ and $\hat{\vb l}_-$ are measured in the $\{\hat{\mathbf{n}}, \hat{\mathbf{k}}, \hat{\mathbf{r}}\}$ coordinate system within their respective quark rest frames.} 
\end{figure}

For 2-qubit systems, a mixed state $\rho$ is separable (non-entangled) if and only if it can be written as $\rho = \sum_{k} p_k \rho^a_k \otimes \rho^b_k$ with $p_k \ge 0$ and $\sum_k p_k = 1$, while an entangled state is defined as non-separable. According to the Peres-Horodecki criterion \cite{Peres:1996dw, Horodecki:1997vt}, in the helicity basis $\{\hat{\mathbf{n}}, \hat{\mathbf{r}}, \hat{\mathbf{k}}\}$ of the center-of-mass frame as shown in Fig.~\ref{fig:frames}, where the coefficients $C_{nr}, C_{rn}, C_{nk}, C_{kn}$, and $B^\pm_i$ vanish, the entanglement can be quantified by the concurrence $\mathcal{C}[\rho] = \max\{\Delta_1/2, \Delta_2/2, 0\}$ \cite{Wootters:1997id}. Here, $\Delta_1 = -1 + C_{nn} + \sqrt{(C_{rr}-C_{kk})^2 + (C_{rk}+C_{kr})^2}$ and $\Delta_2 = -1 - C_{nn} + \sqrt{(C_{rr}+C_{kk})^2 + (C_{rk}-C_{kr})^2}$, with values ranging from $\mathcal{C}[\rho] = 0$ (separable) to $\mathcal{C}[\rho] = 1$ (maximally entangled). When the spin correlation matrix $C_{ij}$ is symmetric ($C_{rk} = C_{kr}$) with $C_{nn} < 0$, the sufficient condition for entanglement simplifies to $\Delta_2 = -1 - C_{nn} + |C_{rr} + C_{kk}| > 0$ \cite{Afik:2020onf, Afik:2022kwm, ATLAS:2023fsd, CMS:2024pts}. For diagonal matrices with nonpositive elements (e.g., $C = \text{diag}(-1, -1, -1)$ for the spin singlet state), this further reduces to the experimentally accessible observable $D = \text{tr}[C]/3 < -1/3$. This quantity can be directly measured as $D = -3\langle\cos\varphi\rangle$, where $\cos\varphi \equiv \hat{\vb l}_+ \cdot \hat{\vb l}_-$, from the angular distribution of decay leptons in top quark pair production near the threshold region, as demonstrated by recent ATLAS and CMS measurements \cite{ATLAS:2023fsd, CMS:2024pts}. For a more detailed description of the concurrence, see the Appendix.

In local hidden variable theories, one form of Bell inequality\cite{Bell:1964kc} is given by
\begin{equation}
\mathcal B =
|\langle a_1 b_1 \rangle+ \langle a_2 b_1 \rangle +
\langle a_1 b_2 \rangle - \langle a_2 b_2 \rangle| \le 2,
\end{equation}
where the values of $a_1, a_2, b_1, b_2$ can either be $1$ or $-1$. Given the density matrix correlation coefficients $C_{ij}$, one can rewrite the expectation value $\langle a_1 b_1 \rangle$ as $\text{tr}[\rho \hat{a}_1 \cdot \vec{\sigma} \otimes \hat{b}_1 \cdot \vec{\sigma}]=\hat{a}_1 \cdot C \cdot \hat{b}_1$ and then express the generalized Bell inequality (CHSH inequality)\cite{Clauser:1969ny, Clauser:1974tg} as
\cite{Fabbrichesi:2021npl}
\begin{equation}
\mathcal B = |
\hat{a}_1 \cdot C \cdot (\hat{b}_1 + \hat{b}_2) + \hat{a}_2 \cdot C \cdot (\hat{b}_1- \hat{b}_2) | \le 2,
\end{equation}
where $\hat{a}_{1, 2}$ and $\hat{b}_{1, 2}$ are the unit vectors along which the quark and antiquark spins can be measured, respectively. Remarkably, it can be shown through algebraic derivation that $\mathcal{B} \leq 2\sqrt{2}$ for a quantum system, and the maximally entangled state can achieve the maximum value of $2\sqrt{2}$. Therefore there exists a range $2<\mathcal B\le 2 \sqrt{2}$ which violates the Bell inequality. The violation of the Bell inequality indicates the presence of Bell nonlocality in quantum mechanics, and it is equivalent to\cite{Cirelson:1980ry, Horodecki:1995nsk}
\begin{equation}
      \mathcal{N} [\rho]\equiv \mu_1+\mu_2-1 >0 , \label{nonlocality}
\end{equation}
where $ \mu_1\ge \mu_2\ge \mu_3 $ are the eigenvalues of the matrix $C^T C$ in descending order. In principle, since all elements of the correlation matrix $C$ can be measured, one can test the Bell nonlocality experimentally through Eq.~(\ref{nonlocality}). For Bell states, which maximally violate the Bell inequality, it gives the nonlocality measure $\mathcal{N}=1$.

\noindent \textit{Quark Pair Productions in photon-gluon fusions.---}
Following the convention used in Refs.~\cite{Uwer:2004vp,Afik:2022dgh, Bernreuther:1993hq,Bernreuther:1997gs,Brandenburg:1998xw,Baumgart:2012ay,
Bernreuther:2015yna,Afik:2022kwm,Afik:2020onf, Aoude:2022imd}, we define the unnormalized spin density matrix for the process $ \gamma^*g \to q  \overline{q} $ as
\begin{equation}
R_{\alpha \alpha', \beta \beta'} = \frac{1}{N} \sum_{\text{color, spin}}  \mathcal M^\ast_{\gamma^*  g \to q^\alpha  \overline{q}^{\alpha^\prime}} \mathcal M_{\gamma^*  g \to q^ \beta  \overline{q}^{\beta^\prime} },
\end{equation}
where $1/N$ represents the average over the intial state color and spin, and $\alpha, \beta, \alpha^\prime, \beta^\prime$ are open spinor indices of quarks and antiquarks produced in the final state.
Using the LO scattering amplitude $\mathcal M_{\gamma^*  g \to q^\alpha  \overline{q}^{\alpha^\prime}}$, it is straightforward to obtain the spin density matrix in an analytical form. Similar to Eq.~(\ref{rho}), since $R$ is a $4 \times 4$ Hermitian matrix, it takes the form
\begin{equation}
       R = \tilde A \mathds 1_4 + \tilde B^+_i \sigma^i \otimes \mathds 1_2
       +\tilde B^-_j \mathds 1_2 \otimes \sigma^j 
       +\tilde C_{ij} \sigma^i \otimes \sigma^j,
\end{equation}
where the overall normalization $\tilde A = \tr R / 4 $ is proportional to the differential cross-section, and the rest of $R$ is related to $\rho$  as follows: $ \rho = R / \tr R = R / (4 \tilde A) $, thus $ B^\pm_i = \tilde B^\pm_i/\tilde A$ and $ C_{ij}=\tilde C_{ij}/\tilde A$. In the above expression for $R$, we have suppressed the spinor indices, which can be recovered as, for example, for the last term as $\tilde C_{ij} \sigma^i_{\alpha \beta} \otimes \sigma^j_{\alpha^\prime \beta^\prime}$ with the first Pauli matrix carrying the quark spinor indices and the second carrying the antiquark spinor indices.

The $R$-matrix and these coefficients can be calculated explicitly 
in the center-of-mass frame of the quark pair with the helicity basis
$ \{\hat{\vb { n}},\hat{\vb {r}},\hat{\vb {k}} \} $
where $ \hat{\vb {k}} $ is the direction of
the outgoing quark, $\hat{\vb {n}}$ is orthogonal to the production plane, 
and $ \hat{\vb {r}} $ is orthogonal to the 
$ \{\hat{\vb {n}},\hat{\vb {k}} \} $ plane as shown in Fig.~\ref{fig:frames}. The indicies $i$ and $j$ of Pauli matrices, which run from $1$ to $3$, are then mapped to the $ \{\hat{\vb { n}},\hat{\vb {r}},\hat{\vb {k}} \} $ coordinates as well. In the center-of-mass frame, let $ p, p', k, k' $ be the 4-momenta of 
the photon, the gluon, the quark, and the antiquark, respectively. The incoming photons are mostly off-shell, thus we write $p^2 = -Q^2 < 0$. The invariant mass $\sqrt{\hat{s}} $ is defined from the Mandelstam variable $\hat{s} = (p+p')^2  = (k+k')^2$,
which is the partonic center-of-mass energy. To make the final expression more compact,
we define three more dimensionless variables: $\alpha =Q^2 / \hat{s}$, $\beta = \sqrt{1-4m^2/ \hat{s}} $, and $z = \hat{\vb p} \cdot \hat{\vb k} $, where $\alpha$ is related to virtuality, $\beta$ is the speed of the produced quark, and $z=\cos \theta$ with $\theta$ being the angle between $\hat{\vb p}$ and $\hat{\vb k}$.

In deep inelastic scattering at EICs, the incoming photons are spacelike virtual photons, which can be either longitudinal or transverse. Typically, the measured unpolarized cross-sections are a linear combination of these two contributions. In EIC experiments, it is potentially possible to separate these two contributions, despite the associated challenges. For example, the incoming photons are predominantly transverse in low-$Q^2$ events, while the longitudinal photons provide the dominant contribution to the concurrence measurements in high $Q^2$ and large $z$ events. Below, we present the cases of longitudinal and transverse photons separately, highlighting their interestingly distinct features. Again, we choose to work in the $ \{\hat{\vb { n}},\hat{\vb {r}},\hat{\vb {k}} \} $ coordinate system, using the physical longitudinal and transverse polarization vectors, $\epsilon_L$ and $\epsilon_T$, respectively, such that $\epsilon \cdot p =0$.

\begin{figure}[H]
        \centering
        \includegraphics[width=0.75\linewidth]{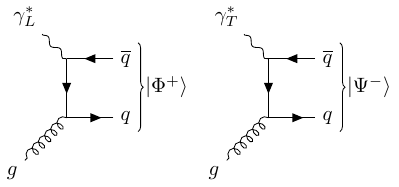}
\caption{\label{fig:feyn} Examples of spin configurations of the longitudinal and transverse photon channels
at $\theta =\frac{\pi}{2}$ ($z = 0$) and the production threshold $\beta = 0$.}
\end{figure}

For the longitudinal polarization contribution, we find that the coefficient
matrix $C$ is given by
\begin{eqnarray}
     &&  C =   \begin{pmatrix} 
                1 & 0 & 0 \\
                0 & -\chi_1& -\chi_2\\
                0 & -\chi_2& \chi_1       
        \end{pmatrix}\\
       \text{with} \quad
      &&   \chi_1 =\frac{1-2z^2+z^2 \beta^2}{1-z^2 \beta^2}, \\ 
     &&    \chi_2 =\frac{2z \sqrt{(1-z^2)(1-\beta^2)}}{1-z^2 \beta^2}.
\end{eqnarray}
Note that $ \chi_1^2+\chi_2^2 = 1$.
In this case, the concurrence $\mathcal C[\rho_L] \equiv 1$ and the nonlocality measure $\mathcal{N} [\rho_L] =1$,
indicating maximal entanglement and maximal violation of Bell inequality. We find that the quark pair in a maximally entangled spin triplet state and the corresponding density matrix is always in the form of a pure state $\rho_L = \ket{\Psi} \bra{\Psi}$, where 
\begin{equation}
        \ket{\Psi} = \frac{1}{2}(\sqrt{1+\chi_1}, \mathrm i \sqrt{1-\chi_1}, 
        \mathrm i \sqrt{1-\chi_1}, \sqrt{1+\chi_1} ). 
\end{equation}
For example as illustrated in Fig.~\ref{fig:feyn}, when $z=0$ ($\theta = \pi/2$), one finds $C=\mathrm{diag}(1,-1,1)$, which indicates $\ket{\Psi} = \ket{\Phi^+}$ and
$\rho_L = \ket{\Phi^+} \bra{\Phi^+}$ 
where $\ket{\Phi^+}= (\ket{\uparrow \uparrow}+\ket{\downarrow \downarrow}) 
/\sqrt{2}$. Interestingly, we also find the same $\rho$ for $\beta =1$ as in the relativistic limit. As detailed in the appendix, the maximal entanglement and pure-state nature of the $q\bar{q}$ pair from the longitudinal channel could be understood through three observations: angular momentum conservation yielding the Bell state $\ket{\Psi^+}$ with $z \sim 1$ in the nonrelativistic limit, helicity conservation leading to the Bell state $\ket{\Phi^+}$ in the high-energy limit, and a Lorentz transformation connecting these regimes.

\begin{figure}[!h]
\centering
\includegraphics[width=0.92\linewidth]{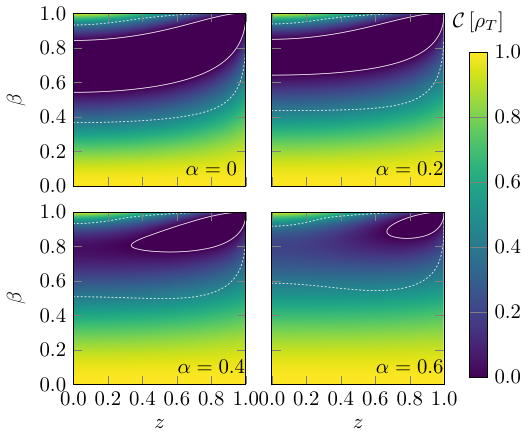}
        \caption{\label{fig:tc} 
        Density plots of the concurrence for the transverse polarized photon contribution at EIC as functions of quark velocity $\beta$ and the scalar projection $z$ at given values of the virtuality parameter $\alpha$, where solid lines and dashed lines indicate the boundaries for entanglement ($\mathcal{C}[\rho_T]=0$) and Bell nonlocality ($\mathcal{N}[\rho_T]=0$), respectively.}
\end{figure}

As to the transverse photon contribution, we find the density matrix is generally in a complicated mixed state. Due to its complexity, the full expression for the transverse contribution, which is presented in the appendix, is used to plot the concurrence $\mathcal{C}[\rho_T]$ in Fig.~\ref{fig:tc} as functions of $\beta$ and $z$ for given values of $\alpha$. It indicates that there is also a wide range of kinematics for observing entanglement and nonlocality.

Regarding the analytic results for the transverse contribution, if the photon is set to be on-shell ($\alpha=0$), then the non-vanishing coefficients of $R$ are:
\begin{eqnarray}
\tilde A &=& F \left[1 + 2\beta^2(1-z^2) -\beta^4(2-2z^2+z^4)\right],\\
        \tilde C_{nn} &=& -F \left[ 1-2\beta^2+\beta^4(2-2z^2+z^4)\right],\\ 
        \tilde C_{rr} &=& -F \left[1-\beta^2(2-\beta^2)(2-2z^2+z^4)\right], \\
        \tilde C_{kk} &=& -F \left[1-2\beta^2 z^2 (1-z^2) - \beta^4(2-2z^2+z^4)\right],\quad \\
        \tilde C_{kr} &=& \tilde C_{rk} =
        2F \beta^2 \sqrt{1-\beta^2} z (1-z^2)^{3 /2}, \\
       && \text{with} \quad F = \frac{e^2_Q g^2_s}{(1-\beta^2 z^2)^2}.
\end{eqnarray}
These coefficients
are very similar to the ones of the process $ g  g \to q  \overline{q} $\cite{Afik:2022kwm,Afik:2020onf, Afik:2022dgh,Baumgart:2012ay,
Bernreuther:2015yna, Aoude:2022imd}, while only the factor $F$ is different. Notably, the matrix $\tilde C$ is not symmetric in the case
of virtual photons ($\alpha > 0$). 

Let us comment on the two asymptotic limits for the transverse contribution to illustrate the relevant physics. At the production threshold, which is equivalent to the nonrelativistic limit by setting $\beta = 0$, the spin correlation is then found to be $C_{ij}=  -\delta_{ij}$ which gives $\rho_T = \ket{\Psi^-} \bra{\Psi^-},$ where $\ket {\Psi^-} $ is the spin singlet state. To understand this result, we note that the amplitude indicates that the transverse photon and gluon form a spin singlet state. In addition, the produced quarks are emitted without angular dependence in an s-wave with zero orbital angular momentum. In this case, conservation of angular momentum requires that the spin of the final state $q\bar{q}$ pair be zero. Meanwhile, in the high energy limit and near $z=0$ region, it is found that $q\bar{q}$ is in the spin triplet $\ket{\Phi^-}$ state due to the orbital angular momentum contribution. As expected, this result is in agreement with that in the $gg\to q\bar{q}$ channel\cite{Afik:2020onf, Afik:2022kwm}.

At the end, to measure the spin configuration of the quark pair, one can utilize the self-analyzing property of weak decays, as the momentum distributions of the decay products are correlated with the spins of the parent particles. Let $\hat{\vb l}_+$ and $\hat{\vb l}_-$ represent the direction vectors of fermions produced in weak decays, measured in the rest frames of their respective parent particles, as illustrated in Fig.~\ref{fig:frames}. The cross-section for producing the lepton pair can be expressed as
\begin{equation}
        \frac{\dd \sigma}{\dd \Omega_+ \dd \Omega_-} 
        \propto \tr\left[(\Gamma_+ \otimes \Gamma_-)R\right], 
\end{equation}
where $\Gamma_{\pm}$ represents the decay density matrices for $q\bar{q}\to l^++l^-+X$ in the rest frames of the parent particles, defined as follows\cite{Baumgart:2012ay}:
\begin{equation}
        \Gamma_{\pm} = \frac{\mathds 1_2+\kappa_{\pm}(\hat{\vb l}_\pm \cdot \bm{\sigma})}{2},
\end{equation} 
where $\kappa_+$ and $\kappa_-$ are the spin-analyzing powers. Since $\Gamma_{\pm}$ depends on the dot product between the Pauli matrices $\sigma^i$ and the lepton's direction, and given that $\tr[\sigma^i] = 0$ and $\tr[\sigma^i \sigma^j] = 2\delta^{ij}$, we can see that the spin information is transferred to the lepton's motion by taking the trace of $(\Gamma_{+} \otimes \Gamma_{-}) R$. Consequently, as the produced $q\bar{q}$ pair is unpolarized with $B^\pm =0$, we can obtain the normalized cross-section
\begin{equation}
        \frac{1}{\sigma}\frac{\dd \sigma}{\dd \Omega_+ \dd \Omega_-} 
        = \frac{1 + \kappa_+ \kappa_- 
\hat{\vb l}_+ \cdot C \cdot \hat{\vb l}_-}{(4\pi)^2}. 
\end{equation}
The correlation coefficients of the matrix $C$ can be determined by measuring the weighted angle averages\cite{Bernreuther:2015yna, Fabbrichesi:2021npl}
\begin{equation}
        C_{ab} =  \frac{9}{\kappa_+ \kappa_-} \; \frac{1}{\sigma} \int_{-1}^1 \dd \xi_{ab} 
        \frac{\dd \sigma}{\dd \xi_{ab}} \xi_{ab} 
        = \frac{9} {\kappa_+ \kappa_-}\langle \xi_{ab}\rangle,
\end{equation}
with the variable $ \xi_{ab} = (\hat{\vb l}_+ \cdot \hat{\vb a}) (\hat{\vb l}_- \cdot \hat{\vb b}) $, 
where the vectors $\hat{\vb a}, \hat{\vb b}$ can be chosen to align with the helicity basis $\{\hat{\vb n}, \hat{\vb r}, \hat{\vb k} \}$. As a special example, the averaged trace $D=\text{tr}C/3$ can be obtained through\cite{ATLAS:2023fsd, CMS:2024pts}
\begin{eqnarray}
 D &=&  \frac{3}{\kappa_+ \kappa_-} \; \frac{1}{\sigma}\int_{-1}^1 \dd \cos \varphi 
        \frac{\dd \sigma}{\dd \cos \varphi} \cos \varphi \notag \\   
   &=& \frac{3}{\kappa_+ \kappa_-} \langle \cos \varphi \rangle.
\end{eqnarray}
where $ \varphi $ is the relative angle 
between $\hat{\vb l}_+$ and $\hat{\vb l}_-$ that $\cos \varphi = \hat{\vb l}_+ \cdot \hat{\vb l}_-$. Possible measurements of $C_{ij}$ can be cast into two categories as follows:
\begin{itemize}
\item Heavy quark pair channels such as $b \overline{b}$ or $c \overline{c} \rightarrow (l^+ \nu q) (l^- \bar{\nu} \bar{q})$: By measuring the lepton pair in the decay product, one can extract the transferred spin correlation information from the heavy quark pair, thereby studying the entanglement and Bell nonlocality at an EIC. (For similar possible decay channels at the LHC, see, e.g., Ref.~\cite{Afik:2024uif}.) 
In this channel, the decayed neutrinos cannot be detected directly, as they appear only as missing momenta in the measurements~\cite{CMS:2024pts, ATLAS:2023fsd}. 
\item  $\Lambda$ and $\bar{\Lambda}$ hyperons channel $s \overline{s} \rightarrow \Lambda \overline{\Lambda} + X
\rightarrow (\pi^- p) + (\pi^+ \overline{p}) + X$\cite{Tornqvist:1980af}. Presumably,
the $\Lambda$ baryons largely inherit their spins from the strange quarks\cite{deFlorian:1997zj, Chen:2023kqw, Chen:2024qvx}. In the subsequent decay,
the proton behaves similarly to the lepton in the previous case. Thus, one can again carry out measurements of the spin correlations of $s\bar{s}$ in the hyperons' rest frame near the production threshold. Recently, the STAR collaboration has clearly demonstrated that this measurement is possible in $pp$ collisions\cite{STAR:2025njp}.
The advantage here is that the momenta of all decay products can be
detected. In turn, measurements of entanglement can help study the spin transfer between the strange quark and the hyperon\cite{Zhao:2024usu}.
\end{itemize}
While our calculations are performed at the parton level, the ATLAS measurement \cite{ATLAS:2023fsd} demonstrated that entanglement survives parton shower and hadronization with only modest dilution (less than $20\%$), indicating that precise EIC measurements can be achieved through comprehensive Monte Carlo simulations that account for parton shower and hadronization effects. Furthermore, to successfully probe quantum entanglement at an EIC, measurements should acquire: (i) precise angular correlation measurements of leptons in heavy quark rest frames or $\Lambda \bar{\Lambda}$ systems, (ii) reliable background modeling and subtraction, and (iii) Monte Carlo-based detector corrections. To study the relevant physics more precisely, it would be ideal to have the experimental capability to separate longitudinal and transverse photon contributions, at least through kinematic suppression.

\noindent \textit{Summary and Outlook.---} To conclude, this work demonstrates that an EIC can offer a unique and clean environment to study quantum entanglement and Bell nonlocality in high-energy collisions. By analyzing spin correlations in quark-antiquark pairs from photon-gluon fusion, we show that maximal entanglement can be achieved, particularly with longitudinally polarized photons, and that there is a large kinematic window for transverse photons. Thus, an EIC enables experimental verification of entanglement and Bell nonlocality in a new regime. This study can help establish the EIC as a novel experimental platform for probing fundamental quantum mechanical phenomena, bridging quantum information theory with high-energy hadronic physics and opening new avenues for testing the foundations of quantum mechanics in previously unexplored kinematic regimes.

Let us also make some brief comments on future outlooks. Similar measurements can be carried out in $eA$ collisions at an EIC. When an entangled quark-antiquark pair traverses the nuclear medium, it undergoes multiple scatterings with the target nucleus, causing decoherence. Thus, concurrence may serve as a quantifiable tool to probe the properties of the nuclear environment. We will leave this study for future work. Lastly, ultra-peripheral collisions at the LHC allow for high-energy quasi-real photon scattering off proton or nuclear targets, similar to what occurs at an EIC. It would be interesting to extend this study to such events if statistics permit.

\noindent \textit{Note added:} After the submission of this manuscript, complementary results were reported by Fucilla and Hatta~\cite{Fucilla:2025kit} for diffractive scattering events.

{\bf Acknowledgments:} This work is supported in part by the Ministry of Science and Technology of China under Grant No. 2024YFA1611004, and by the CUHK-Shenzhen University Development Fund under Grant No. UDF01001859. We thank Alfred Mueller, Feng Yuan, Yoshitaka Hatta, Shu-Yi Wei, Yuxiang Zhao, and Jian Zhou for useful comments.

\bibliography{references}

\appendix
\begin{widetext}
\section{Quantifying Entanglement}

To determine if a mixed state is entangled, one can employ the Peres-Horodecki criterion \cite{Peres:1996dw, Horodecki:1997vt}. It states that if a mixed state
$\rho$ is separable, its partial transpose
\begin{equation}
\rho^{T_2}= \sum_k p_k \rho^a_k \otimes \left( \rho^b_k \right)^T
\end{equation}
is non-negative (i.e., positive semidefinite), since $\left( \rho^b_k \right)^T$ is also Hermitian and thus represents a valid density matrix. Consequently, if any eigenvalue of $\rho^{T_2}$ is negative, the state $\rho$ must be entangled. Moreover, for 2-qubit systems, this condition is both necessary and sufficient for entanglement.

In the helicity basis $\{\hat{\vb n}, \hat{\vb r}, \hat{\vb k}\}$, the coefficients $C_{nr},C_{rn}, C_{nk}, C_{kn}$, $B^\pm_i$ vanish. Therefore, according to the Peres-Horodecki criterion, at least one of the following must be positive for the state to be entangled:
\begin{eqnarray}
 && \Delta_1 =-1+C_{nn}+ \sqrt{(C_{rr}-C_{kk})^2+(C_{rk}+C_{kr})^2}, \quad \\
       && \Delta_2 =-1-C_{nn}+ \sqrt{(C_{rr}+C_{kk})^2+(C_{rk}-C_{kr})^2} . \quad
\end{eqnarray}
The corresponding concurrence is then given by:
\begin{equation}
        \mathcal C [\rho] = \max \left\{\Delta_1 / 2, \Delta_2 /2, 0\right\}.\label{concurrence}
\end{equation}
The concurrence, \(\mathcal{C}[\rho]\), is a key quantity for characterizing entanglement, with values ranging between \(0\) and \(1\). A value of \(\mathcal{C}[\rho] = 0\) corresponds to a separable state, while \(\mathcal{C}[\rho] = 1\) indicates a maximally entangled state. It is formally defined as \cite{Wootters:1997id}: $\mathcal{C}[\rho] = \max \{0, \lambda_1 - \lambda_2 - \lambda_3 - \lambda_4\}$, 
where \(\lambda_i\) are the eigenvalues of the Hermitian matrix $\mathcal{R} = \sqrt{\sqrt{\rho} \tilde \rho \sqrt{\rho}} $ in descending order, and the spin-flipped density matrix is $\tilde \rho = (\sigma^y \otimes \sigma^y) \rho^* (\sigma^y \otimes \sigma^y)$. In the case considered in this work, the expression given in Eq.~(\ref{concurrence}) coincides with the formal definition of the concurrence.  

For a two-qubit system with $B_i^\pm = 0$ (as found in our calculation), if it has maximal entanglement ($\mathcal C[\rho] = 1$), then it must be a pure state. This conclusion can be shown as follows. If $B_i^\pm = 0$, we have $\tilde \rho = \rho$, then the Hermitian matrix $\mathcal R = \rho$. Therefore, the $\lambda_i$'s are the eigenvalues of $\rho$ as well. Since $\mathcal C[\rho] = \lambda_1-\lambda_2-\lambda_3-\lambda_4 = 1$, $\text{tr}\rho =\sum_{i=1}^4 \lambda_i =1$, and $0\le \lambda_i\le 1$, the only solution is $\lambda_1=1$ and $\lambda_2=\lambda_3=\lambda_4=0$. Since $\rho$ has only one non-zero eigenvalue (has a rank of $1$), it is corresponding to a pure state.

\section{Analysis of the Longitudinal Channel}

To understand why the longitudinal polarization leads to maximal entanglement and pure states, let us comment on the spin structure of the produced $q\bar{q}$ pair in both the non-relativistic limit and the high energy limit, then show the Lorentz transformation that connects these two regimes.

\begin{enumerate}
\item In the non-relativistic limit $\beta \to 0$,  we find that the $q\bar{q}$ state is a $p$-wave configuration (with orbital angular momentum $L=1$), and the spin structure corresponds to a Bell triplet state $\ket{\Psi^+} = \frac{1}{\sqrt{2}}( \ket{\uparrow \downarrow}+\ket{\downarrow \uparrow})$, as required by angular momentum conservation along the beam direction. Since
\begin{equation}
       \tilde A = \frac{2\alpha \beta^2(1-z^2) e_Q^2 g_s^2}{(1+\alpha^2)(1-z^2\beta^2)},
\end{equation}
one finds that the scattering cross-section is proportional to$ \left[P^1_1(\cos\theta)\right]^2$, corresponding to a p-wave state. Therefore, there is a nonzero component of orbital angular momentum along the beam direction, which is the same as the gluon spin. As a result of angular momentum conservation and choosing the beam as the direction of the spin basis, the spin state of the produced $q\bar{q}$ is identical to that of the longitudinal virtual photon, $ \ket{s=1, m=0} = \ket{\Psi^+} = \frac{1}{\sqrt{2}}\left(\ket{\uparrow \downarrow}+\ket{\downarrow \uparrow}\right).$ The correlation coefficient matrix corresponding to $\ket{\Psi^+}$ is $C' = \mathrm{diag}(1,1,-1)$, which coincides with our results for the longitudinal correlation matrix after setting $z=1$ ($\theta \to  0$) and $\beta=0$ ($\chi_1 = -1$, $\chi_2 = 0$). 

For a general scattering angle $\theta$, the bases before and after the scattering are different. By applying the rotation transformation $\mathcal R(\theta)$, one obtains the spin correlation $C_{ij} = \mathcal R(\theta)_{ii'} \mathcal R(\theta)_{jj'} C'_{i'j'}$ for the scattering angle $\theta$ in the helicity basis $\{\hat{\vb n},\hat{\vb r},\hat{\vb k}\}$ with
\begin{equation}
        C(\theta) = 
        \begin{pmatrix} 
                1 & 0 & 0\\
                0 & \cos 2\theta & -\sin 2\theta\\
                0 & -\sin 2\theta & -\cos 2\theta
        \end{pmatrix}  \quad \text{where} \quad \mathcal R(\theta) = 
       \begin{pmatrix} 
               1 & 0 & 0\\
               0 & \cos \theta & \sin \theta\\
               0 & -\sin \theta & \cos \theta\\
       \end{pmatrix}.
\end{equation}
Compared with our findings for $\beta =0$, we find indeed $\chi_1 = 1-2z^2 = -\cos 2\theta$, $\chi_2 = 2 z \sqrt{1-z^2} = \sin 2\theta $. For example, let $\theta = \pi /2$, then $C = \mathrm{diag} (1,-1,1)$ corresponding to $\ket{\Phi^+} = \frac{1}{\sqrt{2}}\left( \ket{\uparrow \uparrow} + \ket{\downarrow \downarrow} \right) $. This is consistent with the discussion in our manuscript. It thus follows that, in the nonrelativistic limit, the quark spins appear to be entirely determined by the longitudinal photon, and always form a maximally entangled pure state.

\item In the high-energy limit where the $q\bar{q}$ pair is produced with speed $\beta \to 1$, helicity is approximately conserved. This conservation law requires that the $q\bar{q}$ pair forms a Bell triplet state $\ket{\Phi^+} = \frac{1}{\sqrt{2}}\left(\ket{\uparrow \uparrow} +\ket{\downarrow \downarrow} \right)$ in the helicity basis $\{\hat{\mathbf{n}}, \hat{\mathbf{r}}, \hat{\mathbf{k}}\}$. Specifically, helicity conservation dictates that the final state must be either $q_L \bar{q}_R$ or $q_R \bar{q}_L$, where the subscripts denote left- and right-handed helicities. Since the quarks are produced back-to-back, opposite helicities correspond to spins pointing in the same direction, resulting in a superposition of $\ket{\uparrow \uparrow}$ and $\ket{\downarrow \downarrow}$ in the helicity basis.

\item Between these two limits, there exists a transformation relating them. Let us boost from the center-of mass frame to the quark (or anti-quark) rest frame, the angle between $\hat{\mathbf{k}}$ and the beam direction becomes $\theta'$ due to Lorentz contraction. When boosting in the $\pm \hat{\mathbf{k}}$ direction, the $\hat{\mathbf{k}}$ component shrinks, while the $\hat{\mathbf{r}}$ component remains unchanged. For the new angle $\theta'$,
        \begin{equation}
        \cos \theta' = \frac{z \sqrt{1-\beta^2} }{\sqrt{1-z^2 \beta^2}},\quad \text{and} \quad
        \sin \theta' = \frac{\sqrt{1-z^2}}{\sqrt{1-z^2 \beta^2}}.
\end{equation}

\begin{figure}[H]
        \centering
        \includegraphics{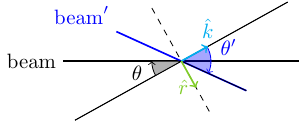}
        \includegraphics{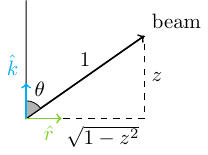}
        \includegraphics{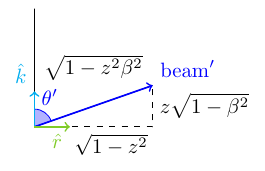}
\end{figure}

Similar to the non-relativistic case, one can again compute the correlation matrix $C$, noting that the only difference is that the rotation angle is $\theta',$ with
\begin{align}
        \cos 2\theta' &= -\frac{1-2z^2+z^2\beta^2}{1-z^2\beta^2} = -\chi_1\\
        \sin 2\theta' &= \frac{2z \sqrt{(1-z^2)(1-\beta^2)} }{1-z^2 \beta^2} 
        = \chi_2
\end{align}
Take the limit $\beta \to 1$, in the (anti) quark's rest frame, the beam is always perpendicular to $\hat{\vb k}$, therefore $\theta' = \pi /2$ and $C = \mathrm{diag}\left( 1,-1,1 \right)$ corresponding to $\ket{\Phi^+} = \frac{1}{\sqrt{2}}\left(\ket{\uparrow \uparrow} + \ket{\downarrow \downarrow}\right)$. The result is consistent with our prediction in part 2 based on helicity conservation, and it confirms that the produced state is a pure Bell state.
\end{enumerate}

Finally, in all the above cases, the spin structure of the produced $q\bar{q}$ pair always forms a maximally entangled pure state. Furthermore, the finding that $\mathcal C[\rho] \equiv 1$ for the longitudinal photon channel implies that the $q\bar{q}$ pair must always be in a pure state, as demonstrated in our earlier analysis in the previous section. We conjecture that the pure state in the longitudinal case appears because the longitudinal photon has only one degree of freedom. In this case, the produced $q\bar{q}$ state is independent of the gluon's helicity, and averaging over two identical pure states still yields a pure state. In contrast, for the transverse case, the interaction is helicity-dependent. Averaging over initial helicities results in a less entangled or mixed state. Therefore, it appears natural that the spin structure of the $q\bar{q}$ pair is always in a maximally entangled pure state.

\section{Density Matrix of the Transverse Channel} 
The density plots in the main text were obtained from the full expression of
the transverse virtual photon contribution, which was omitted for brevity. 
For completeness, we provide the full result here. 
The unnormalized spin density matrix $R$ is a $4 \times 4$ Hermitian matrix
that takes the form
$
       R = \tilde A \mathds 1_4 + \tilde B^+_i \sigma^i \otimes \mathds 1_2
       +\tilde B^-_j \mathds 1_2 \otimes \sigma^j 
       +\tilde C_{ij} \sigma^i \otimes \sigma^j.
$
With the helicity basis
$\{\hat{\vb n},\hat{\vb r}, \hat{\vb k}\}$,
we define the following
variables: 
$\alpha = Q^2/\hat{s}$, 
$\beta = \sqrt{1-4m^2 /\hat{s}}$, 
$z = \hat{\vb p} \cdot \hat{\vb k}$.
The non-vanishing coefficients are
\begin{align}
         \tilde A&= F \left[
         1 + 2 \beta ^2 (1-z^2)
         -\beta ^4 (2-2 z^2+z^4)
         +2 \alpha  (1-\beta ^2) (1+\beta ^2 z^2)
         +\alpha ^2 (1-\beta ^4 z^4)
         \right], \\
         \tilde C_{nn}&= -F \left[
         1 - 2\beta^2 +\beta^4(2-2z^2+z^4)
         + 2 \alpha (1-\beta^2)(1-\beta^2 z^2) 
         + \alpha^2 (1-\beta^2 z^2)^2
         \right],  \\
         \tilde C_{rr}&= -F  \left[
         1-\beta^2(2-\beta^2)(2-2z^2+z^4)+2\alpha(1-\beta^2)(1-\beta^2 z^2)
         +\alpha^2\left(1-2\beta^2(1-z^2+z^4)+\beta^4 z^4\right)
         \right],  \\
         \tilde C_{kk}&= -F \left[
         1 - 2\beta^2 z^2 (1-z^2) - \beta^4(2-2z^2+z^4)+ 2\alpha(1-\beta^2)(1+\beta^2 z^2)
         + \alpha^2\left(1-2\beta^2(1-z^4)-\beta^4 z^4\right)
        \right],\\
          \tilde C_{rk}&= F \left[2\beta \sqrt{(1-\beta^2)(1-z^2)}
          \left(\beta z(1-z^2) 
                +\alpha(1-\beta z) +\alpha^2(1-\beta z^3)\right)
          \right],\\
         \tilde C_{kr}&= F \left[ 2\beta \sqrt{(1-\beta^2)(1-z^2)} 
         \left(\beta z(1-z^2) -\alpha(1+\beta z) -\alpha^2(1+\beta z^3)\right)
         \right], \\
        F &= \frac{e_Q^2 g_s^2}{(1+\alpha)^2(1-\beta^2 z^2)^2}.
\end{align}
If the photon is on-shell ($\alpha=0$), the result reduces to the simplified expression presented in the main text.

\end{widetext}
\end{document}